\documentstyle[12pt]{article}
\catcode`\@=11

\@addtoreset{equation}{section}
\def\theequation{\arabic{section}.\arabic{equation}}

\catcode`\@=11

\def\section{\@startsection{section}{1}{\z@}{3.5ex plus 1ex minus
   .2ex}{2.3ex plus .2ex}{\large\bf}}

\newskip\humongous \humongous=0pt plus 1000pt minus 1000pt

\newif\ifdtup

\def\eqnarray{\let\@currentlabel=\theequation\refstepcounter{equation}
    \global\@eqnswtrue
    \global\@eqcnt\z@\tabskip\@centering\let\\=\@eqncr
    $$\halign to \displaywidth\bgroup\@eqnsel\hskip\@centering
      $\displaystyle\tabskip\z@{##}$&\global\@eqcnt\@ne
       \hfil${{}##{}}$\hfil
      &\global\@eqcnt\tw@ $\displaystyle\tabskip\z@{##}$\hfil
       \tabskip\@centering&\llap{##}\tabskip\z@\cr}
\def\lefteqn#1{\hbox to 4\arraycolsep{$\displaystyle #1$\hss}}
\def\thesection{\arabic{section}.}

\def\appendix{\setcounter{section}{0}
        \def\thesection{Appendix.}
        \def\theequation{\Alph{section}.\arabic{equation}}}

\long\def\@makefntext#1{\parindent 0cm\noindent
\hbox to 1em{\hss$^{\@thefnmark}$}#1}
\def\IR{{\hbox{{\rm I}\kern-.2em\hbox{\rm R}}}}
\def\IH{{\hbox{{\rm I}\kern-.2em\hbox{\rm H}}}}
\def\IC{{\ \hbox{{\rm I}\kern-.6em\hbox{\bf C}}}}
\def\IZ{{\hbox{{\rm Z}\kern-.4em\hbox{\rm Z}}}}
\def\rref#1{(\ref{#1})}
\def\Tr{\hbox{Tr}}
\def\delbar{\,\overline{\mathop{\!\nabla\!}}\,}
\newcommand{\beq}{\begin{equation}}
\newcommand{\eeq}{\end{equation}}
\newcommand{\sint}{{\hskip-.1em\int\!d^2\!x\hskip.1em}}
\newcommand{\CQG}[1]{{\sl Class.~Quant.~Grav.}~{\bf #1}}
\newcommand{\PRD}[1]{{\sl Phys.~Rev.}~{\bf D#1}}

\newcommand{\JMP}[1]{{\sl J.~Math.~Phys.}~{\bf #1}}
\begin{document}
%
%
%
%
\def\citen#1{%
\edef\@tempa{\@ignspaftercomma,#1, \@end, }
\edef\@tempa{\expandafter\@ignendcommas\@tempa\@end}%
\if@filesw \immediate \write \@auxout {\string \citation {\@tempa}}\fi
\@tempcntb\m@ne \let\@h@ld\relax \let\@citea\@empty
\@for \@citeb:=\@tempa\do {\@cmpresscites}%
\@h@ld}
%
\def\@ignspaftercomma#1, {\ifx\@end#1\@empty\else
   #1,\expandafter\@ignspaftercomma\fi}
\def\@ignendcommas,#1,\@end{#1}
%
%
\def\@cmpresscites{%
 \expandafter\let \expandafter\@B@citeB \csname b@\@citeb \endcsname
 \ifx\@B@citeB\relax 
    \@h@ld\@citea\@tempcntb\m@ne{\bf ?}%
    \@warning {Citation `\@citeb ' on page \thepage \space undefined}%
 \else
    \@tempcnta\@tempcntb \advance\@tempcnta\@ne
    \setbox\z@\hbox\bgroup 
    \ifnum\z@<0\@B@citeB \relax
       \egroup \@tempcntb\@B@citeB \relax
       \else \egroup \@tempcntb\m@ne \fi
    \ifnum\@tempcnta=\@tempcntb 
       \ifx\@h@ld\relax 
          \edef \@h@ld{\@citea\@B@citeB}%
       \else 
          \edef\@h@ld{\hbox{--}\penalty\@highpenalty \@B@citeB}%
       \fi
    \else   
       \@h@ld \@citea \@B@citeB \let\@h@ld\relax
 \fi\fi%
 \let\@citea\@citepunct
}
%
\def\@citepunct{,\penalty\@highpenalty\hskip.13em plus.1em minus.1em}%
%
%
\def\@citex[#1]#2{\@cite{\citen{#2}}{#1}}%
%
%
\def\@cite#1#2{\leavevmode\unskip
  \ifnum\lastpenalty=\z@ \penalty\@highpenalty \fi 
  \ [{\multiply\@highpenalty 3 #1
      \if@tempswa,\penalty\@highpenalty\ #2\fi 
    }]\spacefactor\@m}
\let\nocitecount\relax  
%
\begin{titlepage}
\vspace{.5in}
\begin{flushright}
UCD-93-25\\
NSF-ITP-93-115\\
gr-qc/9309002\\
August 1993\\
\end{flushright}
\vspace{.5in}
\begin{center}
{\Large\bf
 Notes on the (2+1)-Dimensional\\[1ex] Wheeler-DeWitt Equation}\\
\vspace{.4in}
{S.~C{\sc arlip}\footnote{\it email: carlip@dirac.ucdavis.edu}\\
       {\small\it Department of Physics}\\
       {\small\it University of California}\\
       {\small\it Davis, CA 95616}\\{\small\it USA}}
\end{center}

\vspace{.5in}
\begin{center}
{\large\bf Abstract}
\end{center}
\vspace{-1ex}
\begin{center}
\begin{minipage}{5in}
{\small
In contrast to other approaches to (2+1)-dimensional quantum gravity,
the Wheeler-DeWitt equation appears to be too complicated to solve
explicitly, even for simple spacetime topologies.  Nevertheless, it is
possible to obtain a good deal of information about solutions and their
interpretation.  In particular, strong evidence is presented that
Wheeler-DeWitt quantization is not equivalent to reduced phase space
quantization.
}
\end{minipage}
\end{center}
\end{titlepage}
\addtocounter{footnote}{-1}

In the continuing search for a realistic quantum theory of gravity, it
has often proven useful to explore simpler models that share the basic
conceptual features of general relativity.  Gravity in 2+1 dimensions
is one such model, providing a fully diffeomorphism-invariant theory
of spacetime geometry that nevertheless avoids many of the technical
difficulties of realistic (3+1)-dimensional gravity.

The goal of this paper is to explore the ramifications of one popular
approach to quantum gravity, the Wheeler-DeWitt equation, in this simple
setting.  We shall see that even with the simplifications of 2+1
dimensions, Wheeler-DeWitt quantization is considerably more complicated
than one might guess, and that at least in its simplest interpretations,
it is not equivalent to other known approaches to quantization.

\section{Canonical Gravity in 2+1 Dimensions}

Before trying to construct a quantum theory, let us briefly review the
canonical formalism for (2+1)-dimensional gravity, as described by
Moncrief \cite{Mon} and Hosoya and Nakao \cite{HosNak}.  We begin with
a spacetime with the topology $\IR\!\times\!\Sigma$, where $\Sigma$ is a
closed surface.  The standard (2+1)-dimensional ADM variables are then
a spatial metric $g_{ij}$ and its canonical momentum $\pi^{ij}$.  By a
classical result of Riemann surface theory, any metric on $\Sigma$ can
be written in the form
\beq
g_{ij} = e^{2\lambda}f^*{\bar g}_{ij} ,
\label{a1}
\eeq
where $f$ is a spatial diffeomorphism and $\bar g_{ij}$ is a ``standard''
metric of constant curvature $1$ (if $\Sigma$ is a sphere), $0$ (if $\Sigma$
is a torus), or $-1$ (if $\Sigma$ is a surface of genus $g>1$).  For a
surface of genus $g>1$, the standard metrics $\bar g_{ij}$ comprise a
$(6g-6)$-dimensional family; for the torus, the family is two-dimensional,
and we can choose
\beq
d\bar s^2 = \tau_2{}^{-1}|dx + \tau dy|^2 ,
\label{a2}
\eeq
where $x$ and $y$ are angular coordinates with period $1$ and $\tau =
\tau_1 + i\tau_2$ is a complex parameter, the modulus.
Corresponding to the decomposition \rref{a1}, the momentum $\pi^{ij}$
can be written as
\beq
\pi^{ij} = e^{-2\lambda}\sqrt{\bar g} \left(
  p^{ij}+ {1\over2}\bar g^{ij}\pi/\sqrt{\bar g}
  + \delbar^iY^j + \delbar^jY^i - \bar g^{ij}\delbar_kY^k \right),
\label{a4}
\eeq
where $\delbar_i$ is the covariant derivative for the connection
compatible with $\bar g_{ij}$, indices are now raised and lowered with
$\bar g_{ij}$, and $p^{ij}$ is a transverse traceless tensor with respect
to $\delbar_i$ (in the language of Riemann surface theory, a
holomorphic quadratic differential).  Roughly speaking, $p^{ij}$ is
canonically conjugate to $\bar g_{ij}$, $\pi$ to $\lambda$, and $Y^i$
to $f$.  More precisely, if we consider a cotangent vector $\delta g_{ij}$
in the space of metrics,
\beq
\delta g_{ij} = \nabla_i\delta\xi_j + \nabla_j\delta\xi_i +
 2e^{2\lambda}\bar g_{ij}\delta\lambda + e^{2\lambda}\delta\bar g_{ij},
\label{a5}
\eeq
the symplectic structure can be read off from the expression
\beq
\int_\Sigma \delta g_{ij} \,\delta \pi^{ij} =
  \int_\Sigma \left\{ \sqrt{\bar g}\delta \bar g_{ij} \,\delta p^{ij}
  + 2\delta\lambda\,\delta\pi
  - 2\,\delta\xi^i\left[ e^{-2\lambda}\sqrt{\bar g}
  \left(\bar\Delta + {1\over2}\bar R \right)\,\delta Y_i
  + {1\over2}\delbar_i\left( e^{-2\lambda}\delta\pi \right)\right]
  \right\} .
\label{a6}
\eeq
(Once again --- and for the remainder of this paper --- indices are
raised and lowered with $\bar g_{ij}$.)  The last term in \rref{a6}
tells us that $\xi^i$ and $Y_i$ are not quite a conjugate pair, a
fact that will cause us considerable grief in the next section.

In terms of the decompositions \rref{a1} and \rref{a5}, the constraints
of canonical gravity become relatively simple.  The momentum constraints
are \cite{Mon}
\beq
\sqrt{\bar g} (\bar\Delta + {1\over2}\bar R) Y_i
  + {1\over2} e^{2\lambda}\delbar_i \left( e^{-2\lambda}\pi\right) = 0
\label{a7}
\eeq
--- note the similarity to the last term of \rref{a6} --- while the
Hamiltonian constraint becomes
\beq
{\cal H} = -{1\over2}{1\over\sqrt{\bar g}}e^{-2\lambda} \pi^2
+ \sqrt{\bar g}e^{-2\lambda}\left[ p^{ij}p_{ij} + 2p_{ij}(PY)^{ij}
  +(PY)_{ij}(PY)^{ij}\right]
+ 2\sqrt{\bar g}\left[ \bar\Delta\lambda - {1\over2}\bar R \right] = 0 ,
\label{a8}
\eeq
where
\beq
(PY)_{ij}
= \delbar_i Y_j + \delbar_jY_i - \bar g_{ij}\delbar_kY^k .
\label{a9}
\eeq
Moncrief has shown that in the classical theory, these constraints fix
$\lambda$ uniquely as a function of $\bar g_{ij}$ and $p^{ij}$, thus
determining a finite-dimensional reduced phase for (2+1)-dimensional
gravity.  A description of dynamics on this reduced phase space depends
on a choice of time slicing; for a foliation by surfaces of constant
mean curvature $\Tr K\!=\!T$, it may be shown \cite{Mon,Carlipobs}
that the Hamiltonian is
\beq
H = \int\!d^2x\,e^{2\lambda} ,
\label{a10}
\eeq
which can in turn be written as a (complicated) function of $\bar g_{ij}$
and $p^{ij}$.

\section{Quantization and the Wheeler-DeWitt Equation}

To quantize this system, we may either solve the constraints classically
and quantize the resulting reduced phase space, or else quantize first and
impose the constraints as conditions to determine physical states.  The
first alternative is discussed in reference \cite{Carlipobs}.  If $\Sigma$
has the topology of a torus, the constraints can be solved explicitly; in
particular, in the York time slicing, $\pi e^{-2\lambda}/\sqrt{\bar g}=T$,
the momentum constraints imply that $Y_i$ vanishes, and the Hamiltonian
constraint has as its solution
\beq
e^{2\lambda} = T^{-1}\left(2p_{ij}p^{ij}\right)^{1/2} .
\label{b1}
\eeq
The system is described by the effective Hamiltonian \rref{a10}, and it
is an easy exercise to find the corresponding Schr\"odinger equation:
\beq
i{\partial\psi\over\partial T} = T^{-1}\Delta_0{}^{1/2}\psi ,
\label{b2}
\eeq
where
\beq
\Delta_0 = -\tau_2{}^2\left( {\partial^2\ \over\partial\tau_1{}^2}
  + {\partial^2\ \over\partial\tau_2{}^2}\right)
\label{b3}
\eeq
is the Laplacian on the torus moduli space with the standard Poincar\'e
metric $d\tau d\bar\tau/\tau_2{}^2$.  The square root in \rref{b2} is
ambiguous, of course; it is customary to take it to be the unique positive
square root, corresponding physically to an expanding universe.

In Wheeler-DeWitt quantization, on the other hand, the Hamiltonian
constraint is not solved classically, but is instead imposed as a
Klein-Gordon-like equation of motion constraining physical states.
In this approach, we no longer have the luxury of choosing a time
slicing, and in particular cannot require that $\Tr K$ be constant on
each slice.  As a consequence, the momentum constraints no longer
require $Y_i$ to vanish, but imply instead that
\beq
Y_i = -{1\over2} \bar\Delta^{-1}
  \left[{1\over\sqrt{\bar g}}e^{2\lambda}
  \delbar_i\left(e^{-2\lambda}\pi\right)\right] .
\label{b4}
\eeq
To obtain the Wheeler-DeWitt equation, we are instructed to insert this
expression into the Hamiltonian constraint and to make the further
substitutions
\beq
\pi = -{i\over2}{\delta\ \over\delta\lambda},\qquad
p^{ij} = -{i\over\sqrt{\bar g}}{\delta\ \over\delta{\bar g}_{ij}} .
\label{b5}
\eeq
For the torus, we find that
\begin{eqnarray}
\Biggl\{
{1\over8}{\delta\ \over\delta\lambda}
  e^{-2\lambda}{\delta\ \over\delta\lambda}
  &+& {1\over2}e^{-2\lambda}\Delta_0 + 2 \bar\Delta\lambda
  - 2e^{-2\lambda}Y^i[\pi]\bar\Delta Y_i[\pi] \\
  &+& 2e^{-2\lambda}\delbar_i\Biggl[\left(2p^{ij}
      + \delbar^iY^j[\pi] + \delbar^jY^i[\pi]
      - \bar g^{ij}\delbar_kY^k[\pi]\right)Y_j[\pi]\Biggr]
  \Biggr\}\Psi[\lambda,\tau]
  = 0 ,\nonumber
\label{b6}
\end{eqnarray}
with $Y_i[\pi]$ given by \rref{b4}.  The operator ordering in the first
term of \rref{b6} is the simplest for which the Hamiltonian constraint
is at least formally Hermitian with respect to the inner product defined
by the ``measure'' $[d\lambda]$.  This is a natural choice, but it is
by no means unique; we shall return briefly to this issue in section 5.
As promised, the $Y_i$ dependence makes equation \rref{b6} complicated
and nonlocal, reflecting the fact that the symplectic structure mixes
$Y_i$ and $\pi$ --- only in the York time slicing do the momentum and
Hamiltonian constraints disentangle.

\section{Naive Schr\"odinger Interpretation}

It is evident from equation \rref{b6} that explicit solutions of the
Wheeler-DeWitt equation will be difficult to find, even in this simple
model.  The fundamental simplifying feature of 2+1 dimensions is that
the physical phase space, parametrized by $\bar g_{ij}$ and $p^{ij}$,
is finite dimensional, effectively reducing quantum field theory to
quantum mechanics.  This simplification is evident in the $p_{ij}p^{ij}$
term of the Hamiltonian constraint, which reduces to a Laplacian on the
two-dimensional moduli space of the torus.  But unlike other approaches
to quantization, the Wheeler-DeWitt equation retains a complicated
dependence on the scale factor $\lambda$, reflecting the fact that the
time slicing has been left arbitrary.

Even without an exact solution to the Wheeler-DeWitt equation, we can
try to extract some useful information.  To do so, however, we must first
decide on an interpretation of the ``wave functions'' $\Psi[\lambda,\tau]$.
This is a highly nontrivial problem: solutions of \rref{b6} do not
come with a ready-made inner product, and the choice of something like
a Hilbert space structure is necessary to give wave functions a
quantum mechanical interpretation.

One natural guess, the ``naive Schr\"odinger interpretation'' \cite{Kuchar},
is that $|\Psi[\lambda,\tau]|^2$ is simply the relative probability of
finding a spatial slice of the universe with scale factor $\lambda$ and
modulus $\tau$.  To compare Wheeler-DeWitt and reduced phase space
quantization, it is more useful to consider the relative probability of
finding a slice with extrinsic curvature $K\!=\!e^{-2\lambda}\pi$ and
modulus $\tau$; this can be obtained from a functional Fourier transform
\beq
\widetilde\Psi[K,\tau] = \int [d\lambda]\,e^{i\sint K e^{2\lambda}}
  \Psi[\lambda,\tau] .
\label{c1}
\eeq
In particular, the wave function on a York time slice $K=T$ is
\beq
\tilde\psi(T,\tau) = \int [d\lambda]\,e^{iT \sint e^{2\lambda}}
  \Psi[\lambda,\tau] ,
\label{c2}
\eeq
giving the relative probability of finding a geometry with modulus $\tau$
on a slice of constant mean curvature $T$.

We can now ask whether this wave function is equivalent to the wave
function \rref{b2} obtained by fixing the same time slicing before
quantization.  Observe first that by \rref{b4},
\beq
\int [d\lambda]\,e^{ iT \sint e^{2\lambda}} Y_i[\pi]\Psi[\lambda,\tau]
  = 0,
\label{c3}
\eeq
since functional integration by parts can be used to move the action of
$\pi$ to the exponential.  The functional Fourier transform of the
Wheeler-DeWitt equation \rref{b6} then reduces to
\beq
\int [d\lambda]\, e^{ iT \sint e^{2\lambda}}
  \int\! d^2x\,N(x)\left\{ -T^2e^{2\lambda} + e^{-2\lambda}\Delta_0
  + 4\bar\Delta\lambda\right\}\Psi[\lambda,\tau] = 0 ,
\label{c4}
\eeq
where $N(x)$ is an arbitrary function. The expression in brackets is
essentially the classical Hamiltonian constraint for the torus, with
$p_{ij}p^{ij}$ replaced by $\Delta_0$.  Now, however, the constraint
need not be obeyed everywhere, but only inside a functional integral.
In particular, there is no reason to suppose that the wave function
$\Psi[\lambda,\tau]$ has its support only on solutions of this constraint.

To obtain further information about $\tilde\psi(T,\tau)$, we can set
$N=1$ in \rref{c4}.  We find that
\beq
iT^2{\partial\psi\over\partial T}(T,\tau)  =
  -\Delta_0 \int [d\lambda]\,\left\{ e^{ iT \sint e^{2\lambda}}
  \int\! d^2x\,e^{-2\lambda} \right\} \Psi[\lambda,\tau] ,
\label{c5}
\eeq
or, differentiating again,
\beq
\left[{\partial\ \over\partial T}T^2{\partial\ \over\partial T}
  + \Delta_0\right] \psi(T,\tau)
  =  \int [d\lambda]\, \left\{ e^{ iT \sint e^{2\lambda}}
  \left[ 1 - \int\!d^2x\, e^{2\lambda}\,\int\!d^2x\, e^{-2\lambda}\,
  \right]\right\}\Delta_0\Psi[\lambda,\tau] .
\label{c6}
\eeq
Up to minor operator ordering ambiguities, the left-hand side of this
expression is the square of the Schr\"odinger equation \rref{b2} obtained
from solving the constraints before quantizing, and is equivalent to
the ``gauge-fixed Wheeler-DeWitt equation'' obtained by Hosoya and Nakao
\cite{HosNak} and Martinec \cite{Martinec}.  The right-hand side thus
measures the departure of the full Wheeler-DeWitt equation from these
alternative approaches to quantization.

In particular, the right-hand side of \rref{c6} vanishes when $\lambda$
is spatially constant, as it is for any classical solution of the
constraints.  As we have seen above, however, there is no reason to expect
the wave function $\Psi[\lambda,\tau]$ to have its support only on such
solutions.  Moreover, the right-hand sides of \rref{c5} and \rref{c6}
depend not only on $\tilde\psi(T,\tau)$, but on the full Wheeler-DeWitt
wave function $\Psi[\lambda,\tau]$, so our would-be Schr\"odinger equation
must be interpreted as one of an infinite family of equations --- somewhat
analogous to the Schwinger-Dyson equations in quantum field theory ---
for functions
\beq
\tilde\psi_n(T,\tau) = \int [d\lambda]\,\left\{e^{iT \sint e^{2\lambda}}
  \int\!d^2x\,e^{2n\lambda}\right\}\Psi[\lambda,\tau] .
\label{c7}
\eeq
While the particular form of equation \rref{c6} depends on a choice of
operator ordering in the Hamiltonian constraint, I have been unable
to find any alternative ordering in which the Wheeler-DeWitt equation
reduces to a single Schr\"odinger- or Klein-Gordon-like equation for
$\tilde\psi(T,\tau)$.

The details of this argument depend on a specific model, but the general
features seem likely to extend to realistic (3+1)-dimensional gravity.
In particular, the trace of the extrinsic curvature in 3+1 dimensions is
canonically conjugate to $\sqrt{g}$, so the wave function at fixed $K$
should take the form
\beq
\tilde\psi_{(3+1)}[T,\tilde g_{ij}]\sim\int[d(\sqrt{g})]e^{iTV[\sqrt{g}]}
   \Psi_{(3+1)}[\sqrt{g},\tilde g_{ij}] ,
\label{c8}
\eeq
where $V = \int\!d^3x\sqrt{g}$ is the volume of a spatial slice and
$\tilde g_{ij}$ symbolizes the remaining metric components.  If we take
two time derivatives of $\tilde\psi_{(3+1)}$, the integrand will be
multiplied by a nonlocal factor $V^2[\sqrt{g}]$.  But just as in our
(2+1)-dimensional model, the action of the local part of the Hamiltonian
constraint can generate at most a single spatial integral in the integrand.
Unless the nonlocal terms in the Hamiltonian constraint conspire to remove
this mismatch in the number of integrations ---  as they fail to do in
(2+1) dimensions --- it would seem difficult to obtain a simple
Klein-Gordon-like equation for the wave function at constant $K$.  The
inequivalence of Dirac and reduced phase space quantization in a theory
with quadratic constraints has been widely studied --- see, for example,
\cite{AshHor,AshStill,Haj,Haj2} --- but I believe this particular feature
is new.

\section{Gauge-Fixing the Inner Product}

The absence of a simple Schr\"odinger or Klein-Gordon equation for
$\tilde\psi(T,\tau)$ should not in itself be viewed as a reason for
rejecting Wheeler-DeWitt quantization.  There are many possible quantum
theories of gravity in 2+1 dimensions \cite{Carsix}, and we do not yet
know how to choose among them.  The naive Schr\"odinger interpretation has
a somewhat more serious problem, however --- there is no reason to expect
the inner product $\langle\tilde\psi(T,\tau)|\tilde\phi(T,\tau)\rangle$
of two states to be conserved, so the relationship to ordinary quantum
mechanics is problematic.  This difficulty can be resolved, at least
formally, by appealing to a suggestion by Woodard \cite{Woodard} for
redefining the inner product.

Woodard's proposal is that all invariances of the action, including
those generated by the Hamiltonian constraint $\cal H$, should be
gauge-fixed in the inner product of two Wheeler-DeWitt wave functions.
For transformations generated by the diffeomorphism constraints, this
procedure is uncontroversial: such transformations take a metric on a
spatial slice $\Sigma$ to a physically equivalent metric, and must
clearly be factored out to avoid overcounting.  The proper treatment
of the group of transformations generated by the Hamiltonian constraint,
on the other hand, is less obvious.  Such transformations mix metrics
on $\Sigma$ with the corresponding canonical momenta --- in the language
of geometric quantization, they do not preserve the polarization --- and
they have no simple interpretation as transformations of the argument of
the wave function.

One argument in favor of Woodard's approach comes from examining wave
functions determined by path integration.  Consider two Hartle-Hawking
wave functions $\Psi_1[g]$ and $\Psi_2[g]$ on a surface $\Sigma$, obtained
from path integrals on manifolds $M_1$ and $M_2$ with boundaries $\partial
M_1\approx\partial M_2\approx\Sigma$.  On $M_1$, say, the action is
invariant under the transformations
\begin{eqnarray}
\delta g_{ij} &=& [\epsilon{\cal H}, g_{ij}] \nonumber\\
\delta \pi^{ij} &=& [\epsilon{\cal H}, \pi^{ij}]
\label{d1}
\end{eqnarray}
only when $\epsilon=0$ on $\Sigma$ \cite{Teit}, so only this subset of
transformations will be gauge-fixed in the path integral for $\Psi_1[g]$.
A similar statement holds for $\Psi_2[g]$.  The inner product, on the
other hand, can be naturally defined as the path integral over the new
manifold $M$ obtained by attaching $M_1$ and $M_2$ along $\Sigma$.  In this
path integral, however, the transformations \rref{d1} must be gauge-fixed
no matter what the value of $\epsilon$ on $\Sigma$; this additional
gauge-fixing leads directly to Woodard's formalism.

To apply this approach to our (2+1)-dimensional model, we begin with
the inner product
\beq
\left\langle\Psi|\Phi\right\rangle
   = \int [d\lambda]\int{d^2\tau\over\tau_2{}^2}\,
   \Psi^*[\lambda,\tau]\Phi[\lambda,\tau] ,
\label{d2}
\eeq
gauge-fixed to the York time slicing
\beq
\chi = \pi - Te^{2\lambda} = 0 .
\label{d3}
\eeq
The path integral over $\lambda$ can be evaluated \`a\ la Faddeev and Popov
by inserting a factor of unity in the form
\beq
1 = \int [d\epsilon]\,\nu^*[\lambda^\epsilon,\tau^\epsilon]
  \delta[\chi^\epsilon]\nu[\lambda^\epsilon,\tau^\epsilon] ,
\label{d4}
\eeq
where $\lambda^\epsilon$, for instance, denotes the transformed value of
$\lambda$,
\beq
\lambda^\epsilon = \lambda - i\left[ \epsilon{\cal H},\lambda \right] .
\label{d5}
\eeq
A simple computation shows that
\beq
\nu[\lambda,\tau] = {\det}^{1/2}\left| \bar\Delta - {1\over2}T^2e^{2\lambda}
   - {1\over2}e^{-2\lambda}\Delta_0 \right|_{\chi=0} ,
\label{d6}
\eeq
up to operator ordering ambiguities coming from the dependence of $\chi$
on the canonical momentum $\pi$.  Note that unlike the more familiar
Faddeev-Popov determinants of gauge theory, $\nu[\lambda,\tau]$ is an
operator, depending explicitly on the Laplacian $\Delta_0$ on moduli
space.

Finding the gauge-fixed form of the inner product \rref{d2} is now
straightforward: it is easy to check that
\beq
\left\langle\Psi|\Phi\right\rangle = \int {d^2\tau\over\tau_2{}^2}
   \hat\psi^*(T,\tau) \hat\phi(T,\tau)
\label{d7}
\eeq
with
\beq
\hat\psi(T,\tau) = \int [d\lambda]  \nu[\lambda,\tau]\,
  e^{iT \sint e^{2\lambda}} \Psi[\lambda,\tau] .
\label{d8}
\eeq
The wave function $\hat\psi(T,\tau)$ is almost identical to the naive
Schr\"odinger wave function $\tilde\psi(T,\tau)$ of the previous section,
differing only by the presence of the determinant $\nu[\lambda,\tau]$.
Now, however, the inner product \rref{d7} is guaranteed to be conserved:
the parameter $T$ merely labels an arbitrary choice in the gauge-fixing
of the original path integral \rref{d2}, and the final result must be
independent of any such choice.

It is natural to ask how the the added factor $\nu[\lambda,\tau]$ affects
the arguments of the previous section.  Unfortunately, this is a hard
question: equation \rref{d6} is complicated enough that I have been unable
to find a useful explicit expression for the determinant.  Observe, for
instance, that the nonlocal terms in the Wheeler-DeWitt equation \rref{b6}
can no longer be neglected.  Moreover, $\nu[\lambda,\tau]$ depends on the
modulus $\tau$, and does not commute with the Laplacian $\Delta_0$ in the
Hamiltonian constraint; to obtain the analog of the would-be Klein-Gordon
equation \rref{c6}, we would have to know this dependence explicitly.

Nevertheless, we can at least obtain some qualitative information about
the determinant $\nu[\lambda,\tau]$ without too much difficulty. Recall
that the naive Schr\"odinger wave function $\tilde\psi(T,\tau)$ failed to
obey a simple Klein-Gordon equation because the Wheeler-DeWitt functional
$\Psi[\lambda,\tau]$ had support away from solutions of the classical
constraints.  It is therefore interesting to note that $\nu[\lambda,\tau]$
is itself peaked around solutions of the constraints.  Indeed, let us
write
\beq
D = -\bar\Delta+{1\over2}T^2e^{2\lambda}+{1\over2}e^{-2\lambda}\Delta_0
  = D_0 + \left(V-V_0\right)
\label{d9}
\eeq
with
\beq
V = {1\over2}T^2e^{2\lambda}+{1\over2}e^{-2\lambda}\Delta_0 \ ,\qquad
V_0 = \int\!d^2x\,V \ ,\qquad D_0 = -\bar\Delta + V_0 .
\label{d10}
\eeq
Note that $V_0$ is always positive, so $D_0$ is invertible.  It is
not hard to show that
\beq
\nu[\lambda,\tau] = \det\!{}^{1/2}\!\left|D\right|
  = \det\!{}^{1/2}\!\left|D_0\right|\,
    \exp\left\{\sum_{n=1}^\infty{(-1)^{n+1}\over 2n}
    \hbox{Tr}\left[D_0^{-1}(V-V_0)\right]^n\right\} .
\label{d11}
\eeq
But $D_0$ has eigenfunctions and eigenvalues
\beq
\left|mn\right\rangle = e^{2\pi i (mx+ny)}\ ,\qquad
\ell_{mn} = {4\pi^2\over\tau_2}|n - m\tau|^2 + V_0 ,
\label{d12}
\eeq
so
\beq
\nu[\lambda,\tau] = \det\!{}^{1/2}\!\left|D_0\right|\,
  \exp\left\{-{1\over4}\sum{1\over\ell_{mn}}{1\over\ell_{pq}}
  \left|\left\langle mn\left|V-V_0\right| pq\right\rangle\right|^2
  + \cdots\right\} ,
\label{d13}
\eeq
which is clearly peaked at $\langle mn|V-V_0| pq\rangle = 0$, i.e., at
spatially constant values of $\lambda$.  Moreover, for $\lambda$ constant,
the derivative
\beq
{\partial\ \over\partial\lambda}\det\!{}^{1/2}\!\left|D_0\right|
  = \left(T^2e^{2\lambda} - e^{-2\lambda}\Delta_0\right)\cdot
 {\partial\ \over\partial V_0}\det\!{}^{1/2}\!\left|D_0\right|
\label{d14}
\eeq
vanishes precisely at the solutions \rref{b1} of the Hamiltonian
constraint, now interpreted as an operator equation on moduli space.
One can further check that in any regularization in which $\Tr D_0^{-1}$
is positive, this extremum is in fact a maximum.

Woodard's proposal thus leads to wave functions at fixed $\Tr K$ that
receive their main contribution from configurations that nearly satisfy
the classical constraints.  In particular, we see from \rref{d13} that
deviations from a constant scale factor are exponentially suppressed.
Note that for $\lambda$ near its classical value, $V_0\sim T\Delta_0^{1/2}$,
so the eigenvalues $\ell_{mn}$ will be smallest --- and the suppression
strongest --- at small values of $T$, i.e., late times.\footnote{In an
expanding (2+1)-dimensional universe, the value of $\Tr K$ on a constant
mean curvature slice decreases from infinity at the initial singularity
to zero in the far future.}  The analog of the would-be Klein-Gordon
equation \rref{c6} for Woodard's wave functions should therefore be closer
to the true Klein-Gordon equation of reduced phase space quantization,
especially at late times.

\section{Next Steps}

Can the results of the last section be made more precise?  Perhaps.  The
two key problems are to understand the determinant $\nu[\lambda,\tau]$ and
the operator ordering in the Hamiltonian constraint $\cal H$.  For any
fixed $\lambda$, $\nu[\lambda,\tau]$ is a modular invariant function on the
torus moduli space, and as such can be expanded in terms of eigenfunctions
of the Laplacian $\Delta_0$.  For instance, it is not hard to show that the
zeta function for the operator $D_0$ is
\beq
\zeta_{D_0}(s) = V_0^{-s} +
  (4\pi^2)^{-s}\sum_{k=0}^\infty {\Gamma(s+k)\over\Gamma(s)\Gamma(k+1)}
  \left(-{V_0\over4\pi^2}\right)^k E(\tau,s+k) ,
\label{e1}
\eeq
where $E(s,k)$ is the Eisenstein series \cite{Kubota,Lang}
\beq
E(\tau,s) = {\sum_{m,n\in\lower1pt\hbox{$\scriptstyle\IZ$}}}^{\!\!\!\prime}
  \,\, {\tau_2^s\over\left|m+\tau n\right|^s} .
\label{e2}
\eeq
(The prime means that the point $m\!=\!n\!=\!0$ is omitted from the sum.)
Such series have been studied quite extensively, and equation \rref{e2} may
yield some useful information.  Elizalde has pointed out that the sum may
be further simplified by an analog of the Chowla-Selberg relation \cite{Eli}.
Similarly, the exponent in \rref{d11} can be expressed in terms of another
Eisenstein series \cite{Lang},
\beq
E_{x,y}(\tau,s) =
  {\sum_{m,n\in\lower1pt\hbox{$\scriptstyle\IZ$}}}^{\!\!\!\prime}
  \,\, e^{2\pi i(mx+ny)} {\tau_2^s\over\left|m+\tau n\right|^s} .
\label{e3}
\eeq
These quantities are complicated, but perhaps not completely intractable.

As for operator ordering, one possible approach is to search for an ordering
such that $[{\cal H}(x),{\cal H}(x')]=0$.  This is a highly nontrivial task,
however, even without the nonlocal term in \rref{b6}.  For example, the
commutator of $\Delta_0$ and $\bar\Delta\lambda$ is a complicated expression
involving single derivatives with respect to the modulus $\tau$, and there
seems to be no simple ordering which eliminates these terms. Once
again, the Wheeler-DeWitt equation proves to be more complex than one might
have anticipated from other approaches to (2+1)-dimensional gravity.

Using this model to investigate other interpretations of the Wheeler-DeWitt
equation would also be of considerable interest.  Wald \cite{Wald} has
recently suggested an intriguing approach to defining a Hilbert space
structure, based on the analogy between the Hamiltonian formulation of
general relativity and that of the relativistic particle.  This approach
requires a solution of the Cauchy problem for the Wheeler-DeWitt equation
\rref{b6}, however, a task made quite difficult by the presence of nonlocal
terms in the Hamiltonian constraint.

\vspace{.1cm}
\begin{flushleft}
\large\bf Acknowledgements
\end{flushleft}
I would like to thank the Institute for Theoretical Physics at Santa
Barbara, where much of this research took place.  I am grateful for
conversations, in person or by electronic mail, with Emilio Elizalde,
Atsushi Higuchi, and Bob Wald.  This work was support in part by the
U.S.\ Department of Energy under grant DE-FG03-91ER40674 and by the
National Science Foundation under grant PHY89-04035.

\end{document}